**PAPER • OPEN ACCESS**

# Evolution of V339 Del (Nova Del 2013) since 0.37 – 75 Days after Discovery

To cite this article: Y Mueangkon et al 2018 J. Phys.: Conf. Ser. **1144** 012014

View the article online for updates and enhancements.

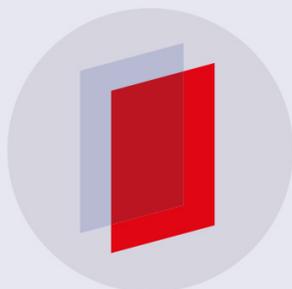

# IOP ebooks™

**Bringing you innovative digital publishing with leading voices to create your essential collection of books in STEM research.**

Start exploring the collection - download the first chapter of every title for free.

# Evolution of V339 Del (Nova Del 2013) since 0.37 – 75 Days after Discovery

Y Mueangkon[1], S Khamrat[1], D Suekong[2], S Aintawiphak[2], A Jaiboe[2], F Surina[2,4,*], M J Darnley[3] and M F Bode[3]

[1] Faculty of Education, Chiang Rai Rajabhat University , 80 Moo 9, Bandu,Muang Chiang Rai 57100, Thailand

[2] Faculty of Science and Technology, Chiang Rai Rajabhat University, 80 Moo 9, Bandu,Muang Chiang Rai 57100, Thailand

[3] Astrophysics Research Institute, Liverpool John Moores University,
Liverpool Science Park, IC2 Building, 146 Brownlow Hill, Liverpool, L3 5RF, UK

[4] National Astronomical Research Institute of Thailand, 260 Moo 4, Donkaew, Maerim, Chiangmai,, 50180, Thailand

*E-mail address: sc_farung@crru.ac.th

**Abstract.** We study the evolution of V339 Del (Nova Del 2013) during 0.37-75 days after discovery. Spectra from the Liverpool Telescope were collected and analysed to find velocity of ejecta ($v_{ej}$), relative flux of radiance with respect to continuum level ($R_\lambda^*$), and *FWHM* of the radiation. The evolution of light curve was explained by adopting an ideal nova light curve as criteria. We found that the evolution of V339 Del during $t = 0.37 – 75$ days can be explained in 7 phases: 1) Initial rise ($t = 0 - 0.6$ days); nova is suddenly brighter from $V \sim 11$ to $\sim 6.4$. A maximum $v_{ej}$ is $\sim 2400$ km/s. $R_\lambda^*$ and *FWHM* first increases and then decreases where this joint ($t = 0.35$ days) turns out to be the first detection of X-ray. 2) Pre-maximum halt ($t = 0.6–1.2$ days); There is a halt of brightness around $V \sim 5.1–5.9$, decreasing $v_{ej}$, increasing $R_\lambda^*$ with decreasing *FWHM*. 3) Final rise (t = 1.2–1.5 days); Nova is brighter again to maximum. The variation of $v_{ej}$ and radiation have similar trend to the halt phase. 4) Maximum ($t = 1.5–2.5$ days); Nova has maximum brightness of $V = 4.45 \pm 0.01$ ($t = 1.67$ days) decreasing $v_{ej}$ and increasing $R_\lambda^*$ until maximum value and the decreasing afterward, while *FWHM* decreases from the final rise. 5) Early decline ($t = 2.5 – 35$ days); Nova has a drop in brightness and $v_{ej}$. The last measurement of $v_{ej}$ is $\sim 1100–1200$ km/s at $t = 35.5$ days. The radiation seems to have 2 distinct phases in this early decline including: First stage ($t = 2.2\sim12$ days) where $R_\lambda^*$ and *FWHM* increase and nebular spectra begins around $t \sim 10$ days. In this stage the nova shell expands optical depth reduces, marking pseudo-photosphere shrink. Second stage ($t = 12\sim35$ days) where $R_\lambda^*$ and *FWHM* decrease and SED shift to near-IR until not visible in optical ($t = 28$ days). Iron curtain (t $\sim$ +25 days) was found near the time of first soft X-ray detection ($t = 35.6$ days). 6) Transition ($t = 35–60$ days); Brightness decreases where $R_\lambda^*$ and *FWHM* gradually increase meaning it reveals deeper pseudo-photosphere. 7) Final decline (t = 60~75 days); Nova is fainter than 6 magnitude from maximum, $R_\lambda^*$ and *FWHM* decrease, Nova is now in nebular phase permanently allowing us to see the surface of white dwarf for the first time.

## 1. Introduction

V339 Del (or Nova Del 2013), a brightest nova outburst in 2013, was discovered at unfiltered CCD magnitude 6.8 by a Japanese amature astronomer Koichhi Itagaki on 2013 Aug. 14.58 UT (i.e. 14:34:14 UT, HJD = 2456519.11163, and designated as $t$ = 0 days). The nova was discovered during its initial rising phase. Optical maximum was on Aug. 16.25 UT when it reached $V \sim 4.3$ [1]. The nova became popular among astronomers due to its brightness.

Nowadays, with panchromatic spectroscopic investigations of each individual nova, people have been tried to explain the mechanism occurred during outbursts of novae. Some novae - e.g. T Pyx for example- were proposed to have two different stages of mass loss with a short-lived phase occurring immediately after outburst, followed by a more steadily evolving and higher mass loss phase [2]. In this work we also noted the physical interpretation of various stages of V339 Del.

## 2. Observations and Data Reduction

We spectroscopically monitored V339 Del at low resolution from $t$ = 0.37 – 75 days after discovery using the 2 m robotic Liverpool Telescope (LT, see [3]). A hundred and twenty spectra of V339 Del including 61 spectra in blue arm and 59 spectra in red arm were secured using the Fibre-fed RObotic Dual-beam Optical Spectrograph (FRODOSpec, see [4]) on the LT over 3900–5700 Å in the blue arm (R $\sim$ 2200) and 5800–9400Å in the red arm (R $\sim$ 2600). Data reduction was performed through an LT pipeline. The *onedspec* package in IRAF[1] was used to analyse all LT spectra.

## 3. Data Analysis

The light curve of V339 Del during $t$ = 0.37 – 75 days from AAVSO was compiled from more than twenty-thousand of $V$ observations whose error < 0.02 magnitude as shown in the top panel of Figure 1. Then LT spectra were used to try to understand what causes the gross changes in the light curve—e.g., changes in mass loss rate from the WD surface as the TNR proceeds—and to derive other parameters of the outburst in future work.

The light curve was compared to ejection velocities ($v_{ej}$) measured from the P Cygni profiles, flux (relatives to continuum level) and FWHM which are measured from $H_\alpha$, $H_\beta$, $H_\gamma$, Fe II (4924, 5018 and 5169 Å) and O I 7775 Å lines as shown in Figure 1 which covers the initial rise phase through the final decline phase of the light curve (t = 0.37-75 days). Considering the modelled temporal evolution of the luminosity ($L_{WD}$), radius ($R_{WD}$) and effective temperature ($T_{eff}$) of the white dwarf (WD) pseudo-photosphere of V339 Del given by [5] together with the X-ray spectroscopic evolution provided in [6] allowed us to determined even more reliable duration of each phases than used to be.

## 4. Results and Discussion

We discuss the spectral evolution based on the idealized nova optical light curve given in [7]. We determined the found that the evolution of V339 Del during $t$ = 0.37 – 75 days can be explained in 7 phases with the as following:

**1) Initial rise ($t$ = 0 - 0.6 days)**; nova was suddenly brighter from $V \sim 11$ (at the pre-nova phase) to $\sim$6.4 [1]. Maximum $v_{ej}$ measured from Balmer lines $\sim$ 1600 – 2200 km/s,  from Fe II lines $\sim$ 1000 - 1900 km/s and  from O I line $\sim$ 2100 - 2400 km/s. Both $R^*_\lambda$ and *FWHM* first increases (during $t$ = 0 - 0.4 days) and then decreases (during $t$ = 0.4 - 0.5 days). We found that the joint between this variation in this phase turns out to be the first detection of X-ray (at $t$ = 0.35 days) given by [6]. They concluded this first X-ray was emitted from the high ionization gas at the centre of WD. At this phase the nova expanded quickly with $R_{WD} = 66 – 84\ R_\odot$, cooled down from $T_{eff} = 10{,}000 – 9{,}000$ K, and $L_{WD} = (1.5\text{-}1.6)\times10^{38}$ erg•s$^{-1}$ as a result [5]. The "*fireball stage* (t = 0 – 5.4 days)" proposed in [5] started here.

---
[1] IRAF is distributed by the National Optical Astronomy Observatory, which is operated by the Association of Universities for Research in Astronomy, Inc., under cooperative agreement with the National Science Foundation.

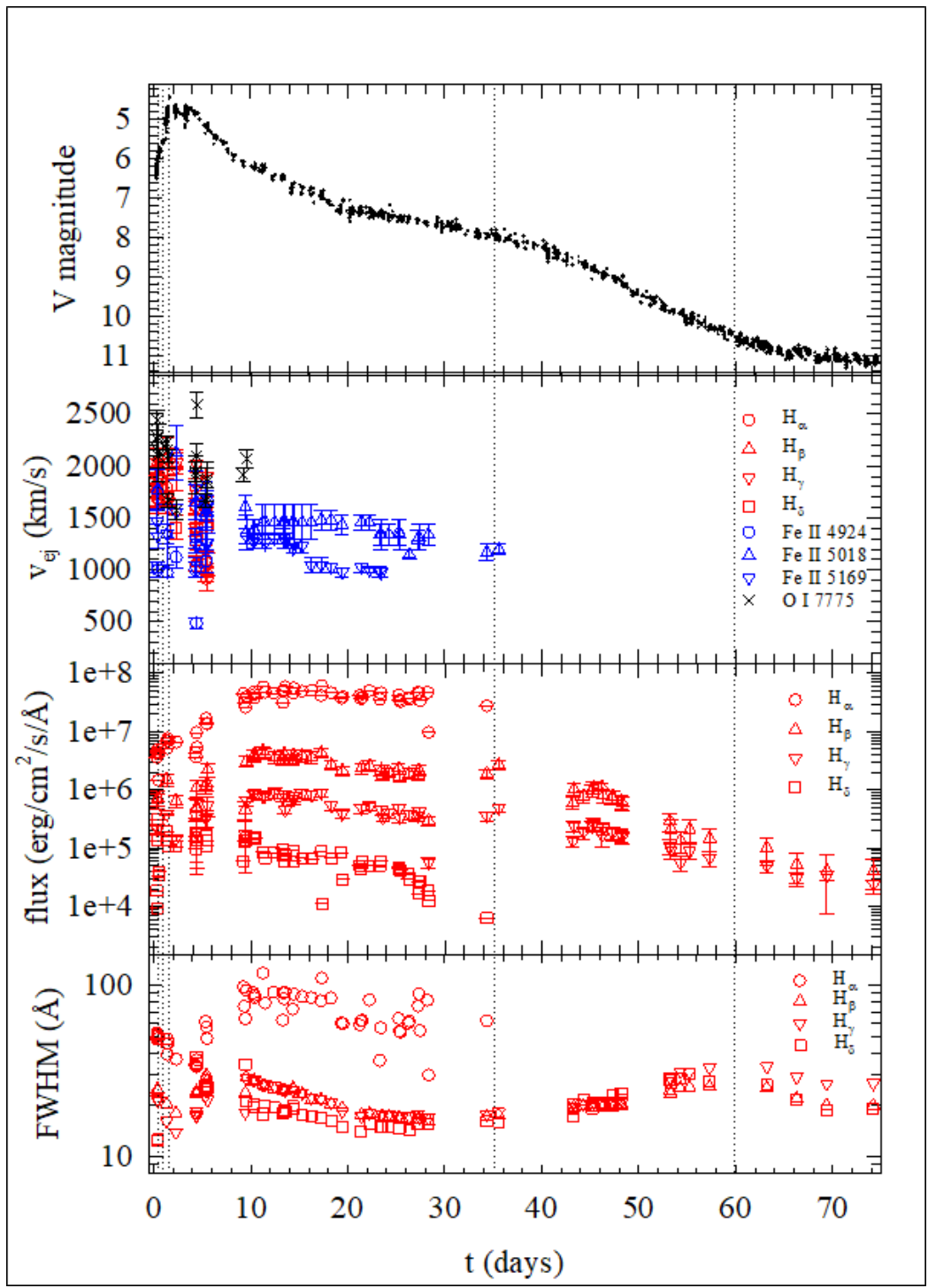

**Figure 1.** Variation in light curve of V339 Del compared to ejection velocities, flux and FWHM from Balmer, Fe II (4924, 5018, and 5169Å), and OI 775 Å lines during *t* = 0.37-75 days. Vertical dotted lines represent 7 phases in the light curve (from the initial rise through the final decline) referred to in the text.

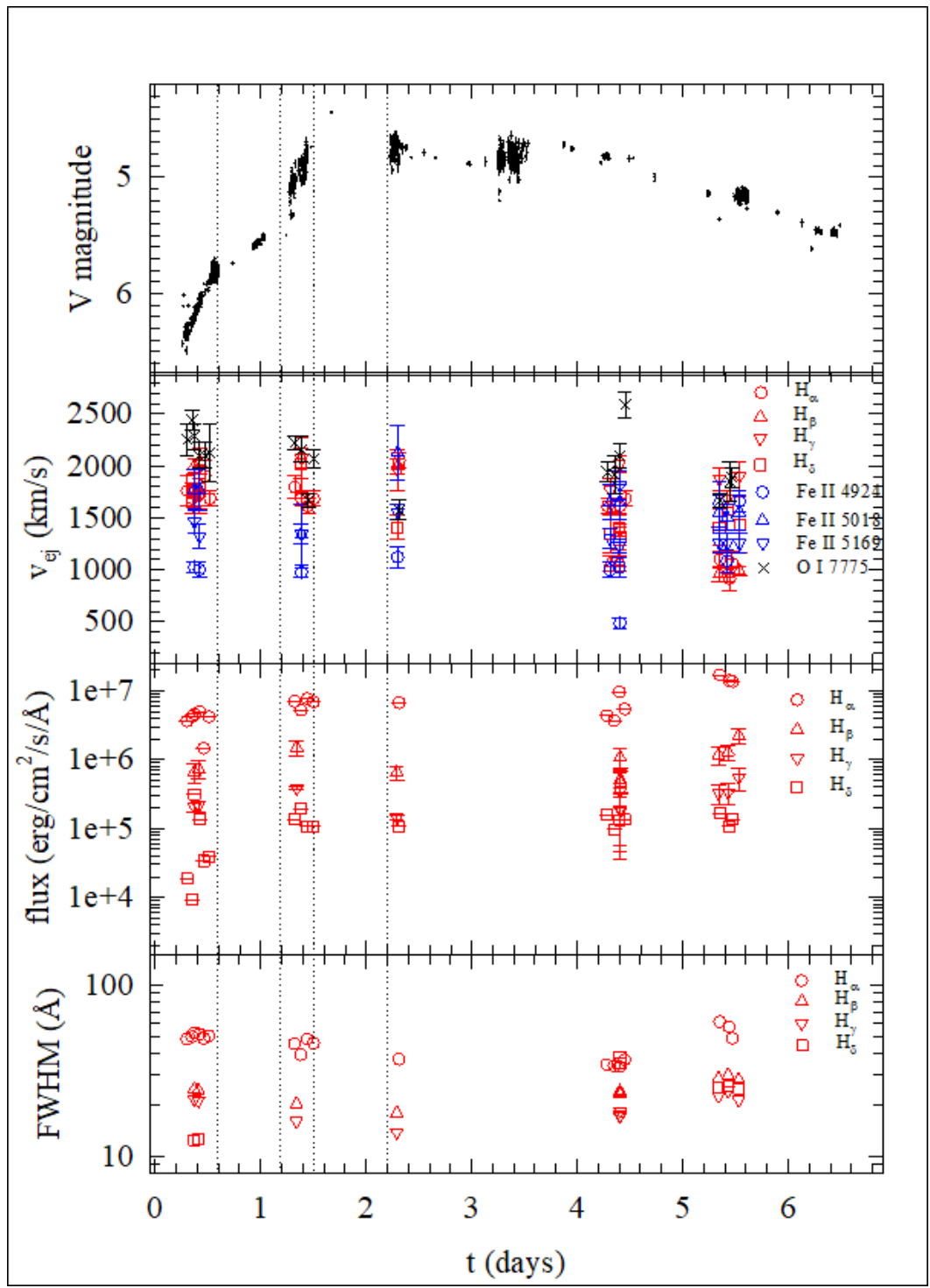

**Figure 2.** Detailed variation in light curve of V339 Del compared to ejection velocities, flux and FWHM from Balmer, Fe II (4924, 5018, and 5169Å), and OI 775 Å lines during $t$ = 0.37-7 days. Vertical dotted lines represent 5 phases in the light curve (from the initial rise through the early decline) referred to in the text.

**2) Pre-maximum halt ($t$ = 0.6–1.2 days)**; There is a halt of brightness around $V \sim$ 5.1–5.9 in AAVSO light curve, decreasing $v_{ej}$, increasing $R_\lambda^*$ with decreasing *FWHM* as shown Figure 2. Our upper limit of this phase (at $t$ = 1.2 days) is corresponding with the date given by [5] mentioning the nova's spectra looked like an late-A type or early F-type star where it began to develop the P Cyg profile. According to the modeled given by [5], $R_{WD}$ was almost unchanged (≤100$R_\odot$), $T_{eff}$ increased from 9,000 – 12,000 K, and $L_{WD}$ increased twice ~ (1.6 – 8.6) × $10^{38}$ erg•s$^{-1}$ in this phase.

**3) Final rise ($t$ = 1.2–1.5 days)**; Nova is brighter again to maximum. The variation of $v_{ej}$ and radiation had similar trend to those in the halt phase.

**4) Maximum ($t$ = 1.5–2.5 days);** Nova has maximum brightness of $V$ = 4.45 ± 0.01 (**$t$ = 1.67 days**), decreasing $v_{ej}$ and increasing $R_\lambda^*$ until maximum value and the decreasing afterward, while *FWHM* keep decreasing. According to the modeled given by [5], $R_{WD}$ expaned quickly until ~300$R_\odot$, causing $T_{eff}$ to cool down to ~ 6,000 K, and $L_{WD}$ decreased to stable at ~ 5 × $10^{38}$ erg•s$^{-1}$.

**5) Early decline ($t$ = 2.5 – 35 days)**; Nova has a drop in brightness ~3.5 magnitude from maximum with decreasing $v_{ej}$. The last measurement of $v_{ej}$ is ~ 1100–1200 km/s at $t$ = 35.5 days. The radiation seems to have 2 distinct phases in this early decline including: First stage ($t$ = 2.2~12 days) where $R_\lambda^*$ and *FWHM* increase and nebular spectra begins around $t \sim$ 10 days. In this stage the nova shell expands optical depth reduces, marking *pseudo-photosphere shrink*. Second stage ($t$ = 12~35 days) where $R_\lambda^*$ and *FWHM* decrease and SED shift to near-IR until not visible in optical ($t$ = 28 days). Iron curtain (t ~ +25 days) was found near the time of first soft X-ray detection ($t$ = 35.6 days).

**6) Transition (t = 35–60 days)**; Brightness decreases where $R_\lambda^*$ and *FWHM* gradually increase meaning it reveals deeper pseudo-photosphere.

**7) Final decline (t = 60~75 days)**; Nova is fainter than 6 magnitude from maximum, $R_\lambda^*$ and *FWHM* decrease, Nova is now in nebular phase permanently allowing us to see the surface of white dwarf for the first time.

**References**

[1]   Munari U, Maitan A, Moretti S and Tomaselli S 2015 *New Astron.* **40** 28
[2]   Surina F, Hounsell R A, Bode M F, Darnley M J, Harman D J and Walter F M 2014 *AJ* **147** 107
[3]   Steele I A et al. 2004  *SPIE* **5489** 679
[4]   Barnsley R M, Smith R J and Steele I A 2012 *Astro Nach.* **333** 101
[5]   Skopal A et al. 2014 *A&A* **569** A112
[6]   Shore S N et al. 2016 *A&A* **590** A123
[7]   Warner B 2008 in *Classical Nova, 2nd Edition* ed. Bode M F and Evans A, Cambridge
         University Press, Cambridge, 16